\RequirePackage[mathlines]{lineno}
\documentclass[%
prd,
superscriptaddress,
nofootinbib,
amsmath,amssymb,
]{revtex4-2}

\usepackage{overpic,graphicx}
\usepackage{dcolumn}
\usepackage{bm}
\usepackage{rotating}
\usepackage{subfigure}
\usepackage{color}
\usepackage[bookmarksnumbered, pdfstartview=FitH,colorlinks,urlcolor=blue, citecolor=blue,linkcolor=blue,] {hyperref}
\usepackage{lineno}
\usepackage{multirow}
\usepackage{amsmath}
\usepackage{makecell}
\begin{document}
\let\oldequation\equation
\let\oldendequation\endequation
\renewenvironment{equation}{\linenomathNonumbers\oldequation}{\oldendequation\endlinenomath}


\title{Track-based alignment for the BESIII CGEM detector in the cosmic-ray test}



\author{A.Q. Guo}
\affiliation{Institute of Modern Physics, Chinese Academy of Sciences, Lanzhou, 730000, China}
\affiliation{University of Chinese Academy of Sciences, Beijing 100049, China}

\author{L.H. Wu}
\email{wulh@ihep.ac.cn}
\affiliation{Institute of High Energy Physics, Chinese Academy of Sciences, Beijing, 100049, China}
\affiliation{University of Chinese Academy of Sciences, Beijing 100049, China}

\author{L.L. Wang}
\email{llwang@ihep.ac.cn}
\affiliation{Institute of High Energy Physics, Chinese Academy of Sciences, Beijing, 100049, China}
\affiliation{University of Chinese Academy of Sciences, Beijing 100049, China}

\author{R.E. Mitchell}
\affiliation{Indiana University, Bloomington, Indiana 47405, USA}

\author{A. Amoroso}
\affiliation{INFN, Sezione di Torino, via P. Giuria 1, 10125 Torino, Italy}
\affiliation{Università di Torino, Dipartimento di Fisica, via P. Giuria 1, 10125 Torino, Italy}

\author{R. Baldini Ferroli}
\affiliation{INFN, Laboratori Nazionali di Frascati, via E. Fermi 40, 00044 Frascati (Roma), Italy}

\author{I. Balossino}
\affiliation{INFN, Sezione di Ferrara, via G. Saragat 1, 44122 Ferrara, Italy}
\affiliation{Institute of High Energy Physics, Chinese Academy of Sciences, Beijing, 100049, China}

\author{M. Bertani}
\affiliation{INFN, Laboratori Nazionali di Frascati,
	via E. Fermi 40, 00044 Frascati (Roma), Italy}

\author{D. Bettoni}
\affiliation{INFN, Sezione di Ferrara,
	via G. Saragat 1, 44122 Ferrara, Italy}

\author{F. Bianchi}
\affiliation{INFN, Sezione di Torino,
	via P. Giuria 1, 10125 Torino, Italy}
\affiliation{Università di Torino, Dipartimento di Fisica, via P. Giuria 1, 10125 Torino, Italy}

\author{A. Bortone}
\affiliation{INFN, Sezione di Torino,
	via P. Giuria 1, 10125 Torino, Italy}
\affiliation{Università di Torino, Dipartimento di Fisica, via P. Giuria 1, 10125 Torino, Italy}



\author{G. Cibinetto}
\affiliation{INFN, Sezione di Ferrara,
	via G. Saragat 1, 44122 Ferrara, Italy}

\author{A. Cotta Ramusino}
\affiliation{INFN, Sezione di Ferrara,
	via G. Saragat 1, 44122 Ferrara, Italy}

\author{F. Cossio}
\affiliation{INFN, Sezione di Torino,
	via P. Giuria 1, 10125 Torino, Italy}

\author{M.Y. Dong}
\affiliation{Institute of High Energy Physics, Chinese Academy of Sciences, Beijing, 100049, China}
\affiliation{University of Chinese Academy of Sciences, Beijing 100049, China}

\author{M. Da Rocha Rolo}
\affiliation{INFN, Sezione di Torino,
	via P. Giuria 1, 10125 Torino, Italy}

\author{F. De Mori}
\affiliation{INFN, Sezione di Torino,
	via P. Giuria 1, 10125 Torino, Italy}
\affiliation{Università di Torino, Dipartimento di Fisica, via P. Giuria 1, 10125 Torino, Italy}

\author{M. Destefanis}
\affiliation{INFN, Sezione di Torino,
	via P. Giuria 1, 10125 Torino, Italy}
\affiliation{Università di Torino, Dipartimento di Fisica, via P. Giuria 1, 10125 Torino, Italy}

\author{J. Dong}
\affiliation{Institute of High Energy Physics, Chinese Academy of Sciences, Beijing, 100049, China}

\author{F. Evangelisti}
\affiliation{INFN, Sezione di Ferrara,
	via G. Saragat 1, 44122 Ferrara, Italy}
\affiliation{Università di Ferrara, Dipartimento di Fisica e Scienze della Terra, via G. Saragat 1, 44122 Ferrara, Italy}

\author{R. Farinelli}
\affiliation{INFN, Sezione di Ferrara, via G. Saragat 1, 44122 Ferrara, Italy}

\author{L. Fava}
\affiliation{INFN, Sezione di Torino,
	via P. Giuria 1, 10125 Torino, Italy}
\affiliation{Università di Torino, Dipartimento di Fisica, via P. Giuria 1, 10125 Torino, Italy}

\author{G. Felici}
\affiliation{INFN, Laboratori Nazionali di Frascati,
	via E. Fermi 40, 00044 Frascati (Roma), Italy}

\author{I. Garzia}
\affiliation{INFN, Sezione di Ferrara,
	via G. Saragat 1, 44122 Ferrara, Italy}
\affiliation{Università di Ferrara, Dipartimento di Fisica e Scienze della Terra, via G. Saragat 1, 44122 Ferrara, Italy}

\author{M. Gatta}
\affiliation{INFN, Laboratori Nazionali di Frascati,
	via E. Fermi 40, 00044 Frascati (Roma), Italy}

\author{G. Giraudo}
\affiliation{INFN, Sezione di Torino,
	via P. Giuria 1, 10125 Torino, Italy}

\author{S. Gramigna}
\affiliation{INFN, Sezione di Ferrara,
	via G. Saragat 1, 44122 Ferrara, Italy}
\affiliation{Università di Ferrara, Dipartimento di Fisica e Scienze della Terra, via G. Saragat 1, 44122 Ferrara, Italy}

\author{S. Garbolino}
\affiliation{INFN, Sezione di Torino,
	via P. Giuria 1, 10125 Torino, Italy}

\author{M. Greco}
\affiliation{INFN, Sezione di Torino,
	via P. Giuria 1, 10125 Torino, Italy}
\affiliation{Università di Torino, Dipartimento di Fisica, via P. Giuria 1, 10125 Torino, Italy}

\author{Z. Huang}
\affiliation{Peking University,
	Beijing 100871, People’s Republic of China}

\author{Y.R. Hou}
\affiliation{University of Chinese Academy of Sciences, Beijing 100049, China}

\author{W. Imoehl}
\affiliation{Indiana University,
	Bloomington, Indiana 47405, USA}

\author{L. Lavezzi}
\affiliation{INFN, Sezione di Torino,
	via P. Giuria 1, 10125 Torino, Italy}
\affiliation{Università di Torino, Dipartimento di Fisica, via P. Giuria 1, 10125 Torino, Italy}

\author{X.L. Lu}
\affiliation{Institute of High Energy Physics, Chinese Academy of Sciences, Beijing, 100049, China}
\affiliation{University of Chinese Academy of Sciences, Beijing 100049, China}

\author{M. Maggiora}
\affiliation{INFN, Sezione di Torino,
	via P. Giuria 1, 10125 Torino, Italy}
\affiliation{Università di Torino, Dipartimento di Fisica, via P. Giuria 1, 10125 Torino, Italy}

\author{F. M. Melendi}
\affiliation{Università di Ferrara, Dipartimento di Fisica e Scienze della Terra, via G. Saragat 1, 44122 Ferrara, Italy}
\affiliation{INFN, Sezione di Ferrara, via G. Saragat 1, 44122 Ferrara, Italy}

\author{R. Malaguti}
\affiliation{INFN, Sezione di Ferrara,
	via G. Saragat 1, 44122 Ferrara, Italy}

\author{A. Mangoni}
\affiliation{INFN, Sezione di Perugia,
	via A. Pascoli, 06123 Perugia, Italy}
\affiliation{Università di Perugia, Dipartimento di Fisica e Geologia, via A. Pascoli, 06123 Perugia, Italy}

\author{S. Marcello}
\affiliation{INFN, Sezione di Torino,
	via P. Giuria 1, 10125 Torino, Italy}
\affiliation{Università di Torino, Dipartimento di Fisica, via P. Giuria 1, 10125 Torino, Italy}

\author{M. Melchiorri}
\affiliation{INFN, Sezione di Ferrara,
	via G. Saragat 1, 44122 Ferrara, Italy}

\author{G. Mezzadri}
\affiliation{INFN, Sezione di Ferrara,
	via G. Saragat 1, 44122 Ferrara, Italy}
\affiliation{Institute of High Energy Physics, Chinese Academy of Sciences, Beijing, 100049, China}

\author{Q. Ouyang}
\affiliation{Institute of High Energy Physics, Chinese Academy of Sciences, Beijing, 100049, China}
\affiliation{University of Chinese Academy of Sciences, Beijing 100049, China}


\author{S. Pacetti}
\affiliation{INFN, Sezione di Perugia,
	via A. Pascoli, 06123 Perugia, Italy}
\affiliation{Università di Perugia, Dipartimento di Fisica e Geologia, via A. Pascoli, 06123 Perugia, Italy}

\author{P. Patteri}
\affiliation{INFN, Laboratori Nazionali di Frascati,
	via E. Fermi 40, 00044 Frascati (Roma), Italy}


\author{A. Rivetti}
\affiliation{INFN, Sezione di Torino,
	via P. Giuria 1, 10125 Torino, Italy}

\author{R.S. Shi}
\affiliation{Institute of High Energy Physics, Chinese Academy of Sciences, Beijing, 100049, China}
\affiliation{University of Chinese Academy of Sciences, Beijing 100049, China}

\author{M. Scodeggio}
\affiliation{INFN, Sezione di Ferrara,
	via G. Saragat 1, 44122 Ferrara, Italy}
\affiliation{Università di Ferrara, Dipartimento di Fisica e Scienze della Terra, via G. Saragat 1, 44122 Ferrara, Italy}

\author{S. Sosio}
\affiliation{INFN, Sezione di Torino,
	via P. Giuria 1, 10125 Torino, Italy}
\affiliation{Università di Torino, Dipartimento di Fisica, via P. Giuria 1, 10125 Torino, Italy}

\author{S. Spataro}
\affiliation{INFN, Sezione di Torino,
	via P. Giuria 1, 10125 Torino, Italy}
\affiliation{Università di Torino, Dipartimento di Fisica, via P. Giuria 1, 10125 Torino, Italy}

\author{B.L. Wang}
\affiliation{University of Chinese Academy of Sciences, Beijing 100049, China}

\author{H.P. Wang}
\affiliation{Institute of High Energy Physics, Chinese Academy of Sciences, Beijing, 100049, China}
\affiliation{University of Chinese Academy of Sciences, Beijing 100049, China}

\author{J.Y. Zhao}
\affiliation{Institute of High Energy Physics, Chinese Academy of Sciences, Beijing, 100049, China}
\affiliation{University of Chinese Academy of Sciences, Beijing 100049, China}

\date{\today}


\begin{abstract}
The Beijing Electron Spectrometer III (BESIII) is a multipurpose detector operating on the Beijing Electron Positron Collider II (BEPCII). After more than ten year's operation, the efficiency of the inner layers of the Main Drift Chamber (MDC) decreased significantly. To solve this issue, the BESIII collaboration is planning to replace the inner part of the MDC with three layers of Cylindrical triple Gas Electron Multipliers (CGEM).

The transverse plane spatial resolution of CGEM is required to be 120 $\mu$m or better. To meet this goal, a careful calibration of the detector is necessary to fully exploit the potential of the CGEM detector. In all the calibrations, the detector alignment plays an important role to improve the detector precision. The track-based alignment for the CGEM detector with the Millepede algorithm is implemented to reduce the uncertainties of the hit position measurement. Using the cosmic-ray data taken in 2020 with the two layers setup, the displacement and rotation of the outer layer with respect to the inner layer is determined by a simultaneous fit applied to more than 160000 tracks. A good alignment precision has been achieved that guarantees the design request could be satisfied in the future. A further alignment is going to be performed using the combined information of tracks from cosmic-ray and collisions after the CGEM is installed into the BESIII detector.
    
\end{abstract}

\maketitle

\tableofcontents

\section{Introduction}
The BESIII spectrometer~\cite{BESIII:2009fln} and the BEPCII $e^+e^-$ collider is a unique machine operating in the $\tau$-Charm region and has collected world record statistics from 2.0 to 5.0 GeV in the last decade. 
This data set yields fruitful physics results and its lifespan is likely be extended for another 5 to 10 years with some necessary upgrades both on the accelerator and detector aspects.  Since the Inner Tracker (IT) has a serious aging effect due to the radiation from beam-related background, replacing the old IT with CGEM detector~\cite{Balossino:2022ywn} is one of the essential upgrades from the detector side. The new IT has a significant radiation hardness, high rate capability, and excellent resolutions both in the longitudinal and transverse directions and it is crucial for the long-term operation of the BESIII detector. 

\begin{figure}[htbp]
\centering
\includegraphics[width=8cm]{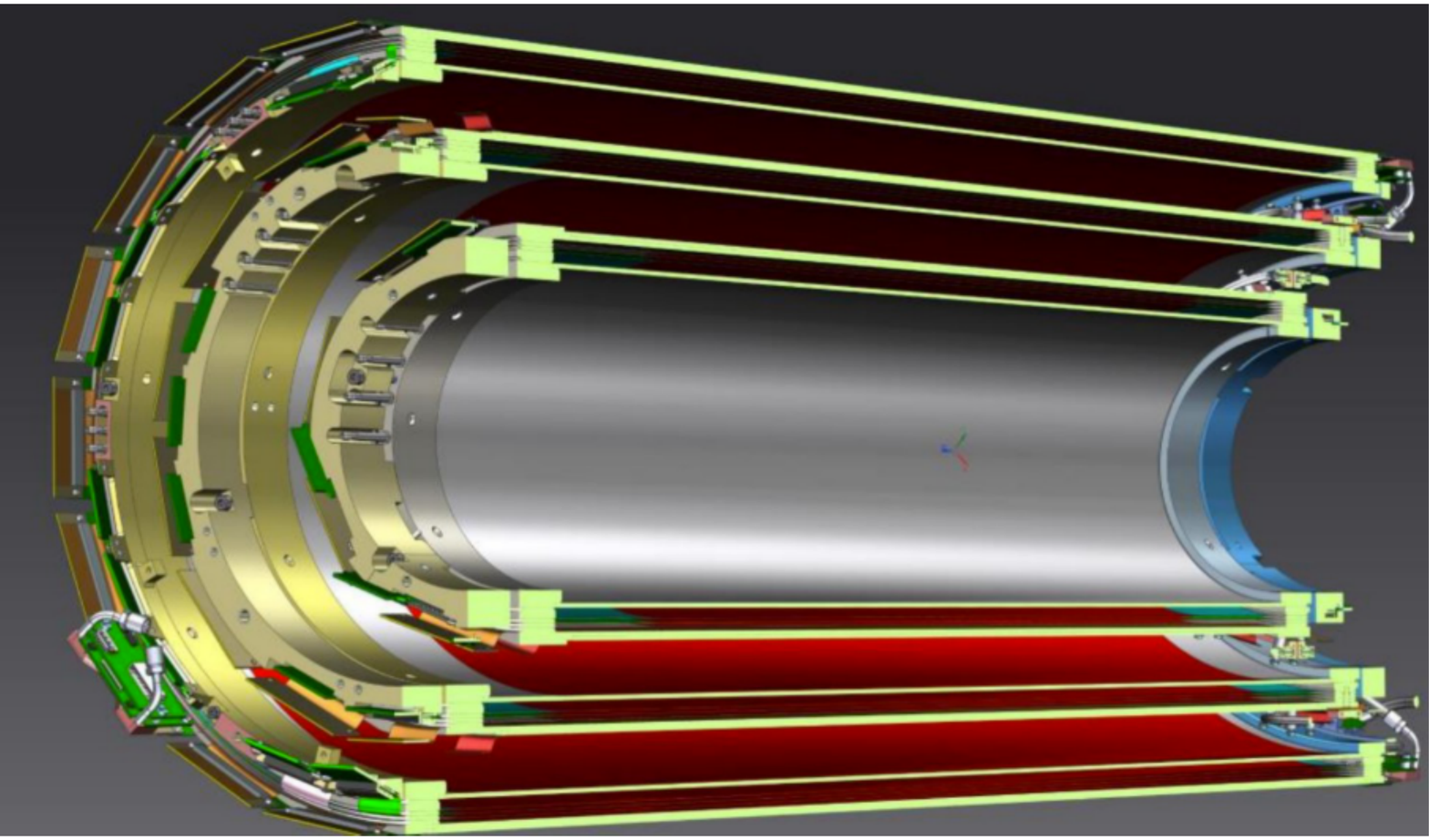}
\includegraphics[width=8cm]{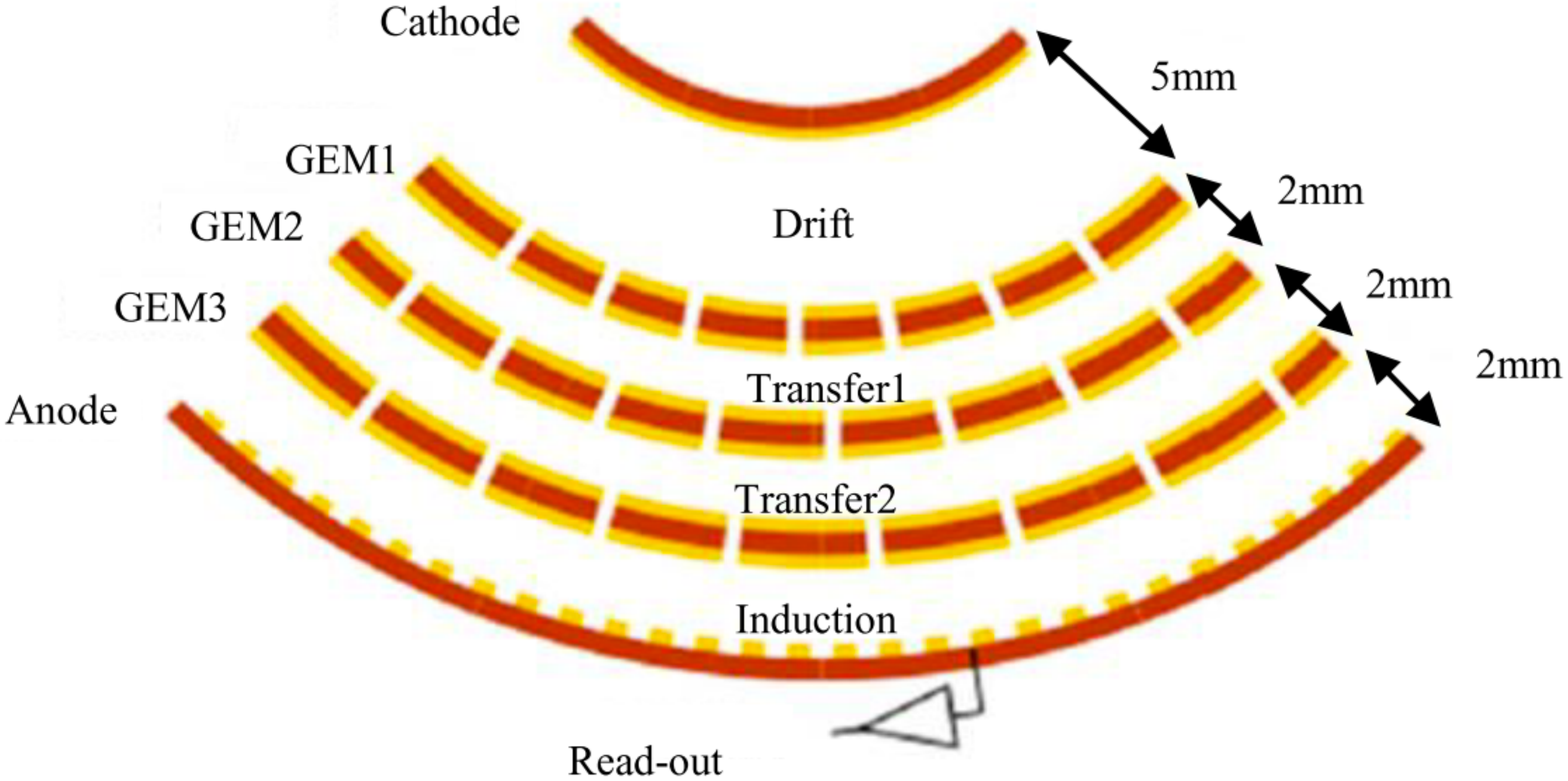}
\caption{Left: the geometry of CGEM detector. Right: the structure of each CGEM layer}
\label{fig::cgem}
\end{figure}

The new IT is composed of 3 layers of CGEM and its geometry can be found in Fig.~\ref{fig::cgem} (Left). Each layer is a triple-GEM composed of concentric cylindrical electrodes (Fig.~\ref{fig::cgem} (Right)): the cathode, three GEM foils, and the readout anode.  Each layer plays as an independent tracking detector providing a 3-D reconstruction of space points along the track with a 2-D readout.  Such a detector will match the requirements for radial resolution ($\sigma_{xy}\sim120\mu m$) of the existing drift chamber and will improve significantly the spatial resolution along the beam direction ($\sigma_{z}\sim220 \mu m$) with very small material budget (less than 1.5\% of $\rm X_0$)~\cite{Amoroso:2016vda}. The new IT will be installed in the available space left by the removal of the inner MDC. The new tracker will achieve the requirements for momentum resolution of $\sigma_{pt}/p_t\sim0.5\%$ at 1\,GeV/c in a 1T magnetic field. 

\section{Detector setup for the cosmic-ray test}

\begin{figure}[htbp]
\centering
\includegraphics[height=7cm]{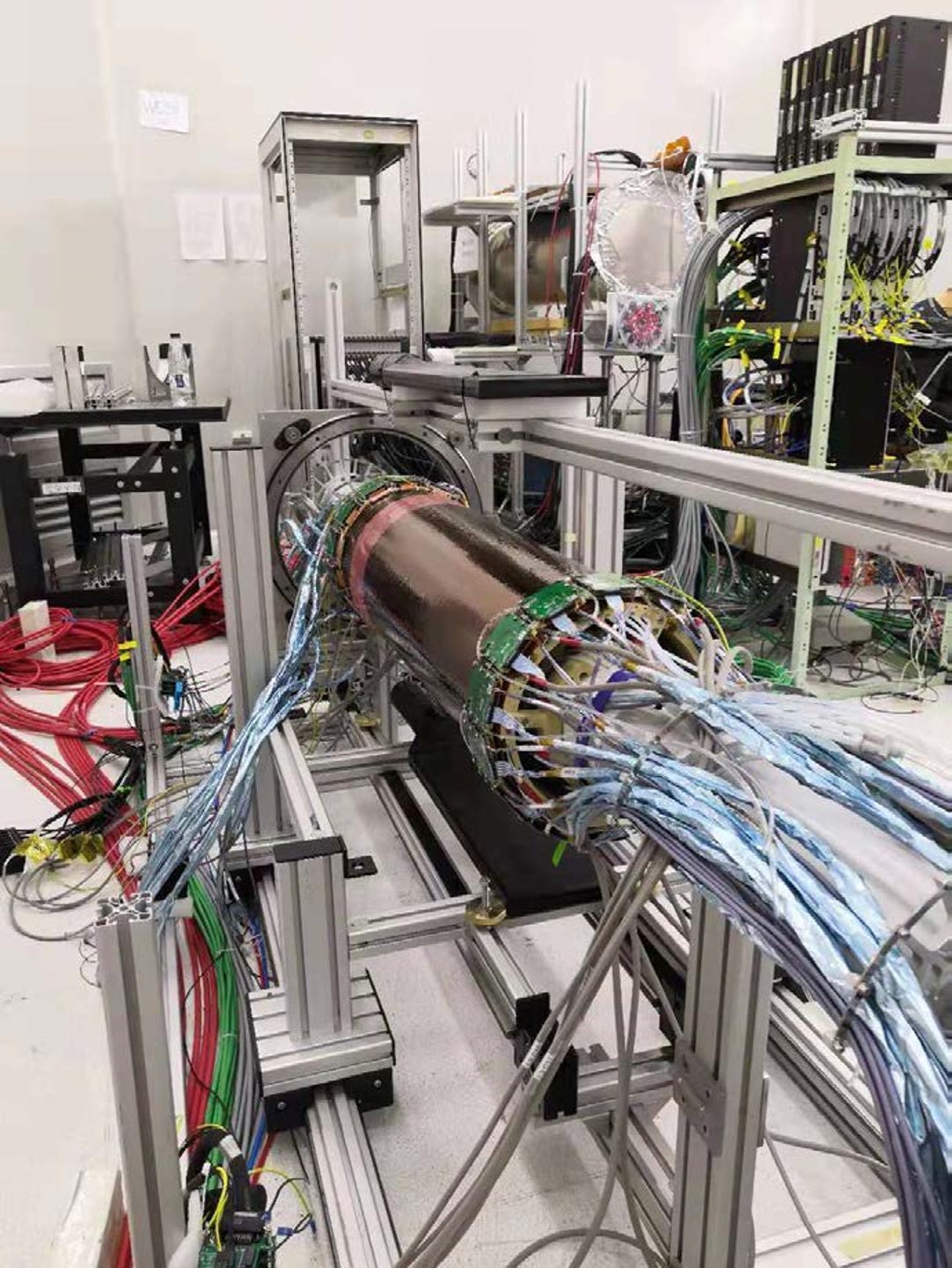}
\includegraphics[width=10cm]{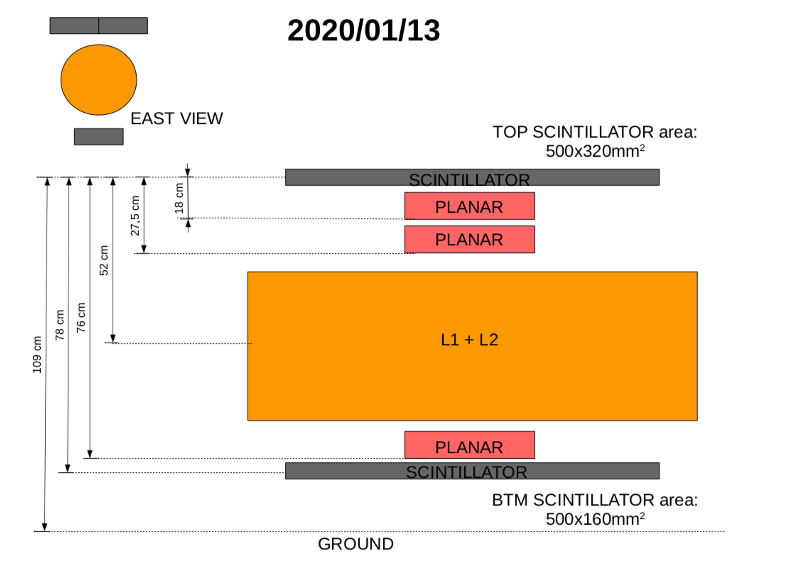}
\caption{Left: the setup of the cosmic ray test for the CGEM. Right: a scheme of the setup with two CGEM (orange box) and the scintillating bars (light dark boxes). }
\label{fig::CosRay_Test}
\end{figure}

Before the installation of CGEM  into the BESIII detector, two of the three CGEM layers have been assembled in the experimental hall of the Institute of High Energy Physics (IHEP) in Beijing. The whole system, including the trigger,  detector, the readout system, is tested by the cosmic ray as shown in Fig.~\ref{fig::CosRay_Test}. 

Two layers of scintillator bars are placed above and below the CGEM detector and serve as a trigger system. 
The CGEM layers have an inner radius of 76.9\,$mm$ and 121.4\,$mm$, while their length is 532\,$mm$ and 690\,$mm$ in the active area. The readout layer is segmented by longitudinal (X) and stereo (V) strips, the length of the strips is fixed for the longitudinal ones, while for the stereo strips it varies from a few millimeters up to tens of centimeters.   The pitch size of the strips is 650\,$\mu m$, therefore the total number of channels instrumented at present is about five thousand. 
The gas mixture used inside the CGEM is $\rm{Ar + iC_{4}H_{10}}$ (90:10) to provide about 55 electrons per m.i.p. in 5 mm. The HV on the electrodes is 835\,V which corresponds to 12 thousand detector gain, while the electric fields between the electrodes are set to 1.5/3/3/5 kV/cm~\cite{zjy:2019}. 
 

The Torino Integrated Gem Electronics for Readout electronics~\cite{Amoroso:2021glb} is used to measure the time and the collected charge. Two ASICs are installed on a Front-End Board (FEB) to read the signal from 128 strips. The chip has two separate branches to extract time and charge information. Every signal crossing threshold on both branches is digitized and transmitted to the off-detector electronics. The charge can be measured with two methods: Sample and Hold mode and Time-Over-Threshold mode~\cite{BESIIICGEM-ITworkinggroup:2020ozx}. The former one was adopted during the cosmic-ray data taking. Once the trigger is sent to the readout chain, the data measured by each chip is collected. The GEM Read Out Card (GEMROC) manages the low voltage of the FEB, the chip configuration, and the data collection. 
Since only two layers were used during the cosmic-ray data taking, the readout electronics have 44 FEB, 11 GEMROC, and two data concentrator cards to send the whole CGEM-IT outputs to the BESIII DAQ system. 

\section{Alignment software}
The BESIII Offline Software System (BOSS) is used to manage the geometry description of the detector, the digitization for the simulation, and the cluster reconstruction for the data. Within the BOSS framework, the measurements from the CGEM-IT will be used to create the 1D cluster and 2D cluster first, then a track-finding algorithm is applied to all the 2D clusters to choose the best candidate on each layer. The selected clusters are fitted by a straight line with the least-squares method. The cluster and the track reconstruction procedures will be explained in the following sections.


\subsection{Cluster reconstruction}
Contiguous firing strips are clustered to form the 1D cluster. Their charge and time information is used to characterize the signal. There are two methods to reconstruct the position: the Charge Centroid and the micro-Time Projection Chamber~\cite{Alexeev:2019rng}. The first averages the position of each strip of the cluster with its charge. The performance of this method is robust, especially for tracks with small incident angle. Meanwhile, the second associate each strip with a bi-dimensional point and it uses a linear fit on the points to extrapolate the position. This method will give better resolution for tracks with large incident angle. A combined position reconstruction method is being developed to take advantage of the two methods above. The Charge Centroid method is used in this study. The clusterization is performed on longitudinal and stereo strips.  The combination of the two measurements gives a 2D cluster. An event display for both 1D and 2D clusters is shown in Fig.~\ref{fig::cluster} top-right subplot.


\subsection{Track finding and fitting for cosmic-ray}
\begin{figure}[htbp]
\centering
\includegraphics[width=7cm]{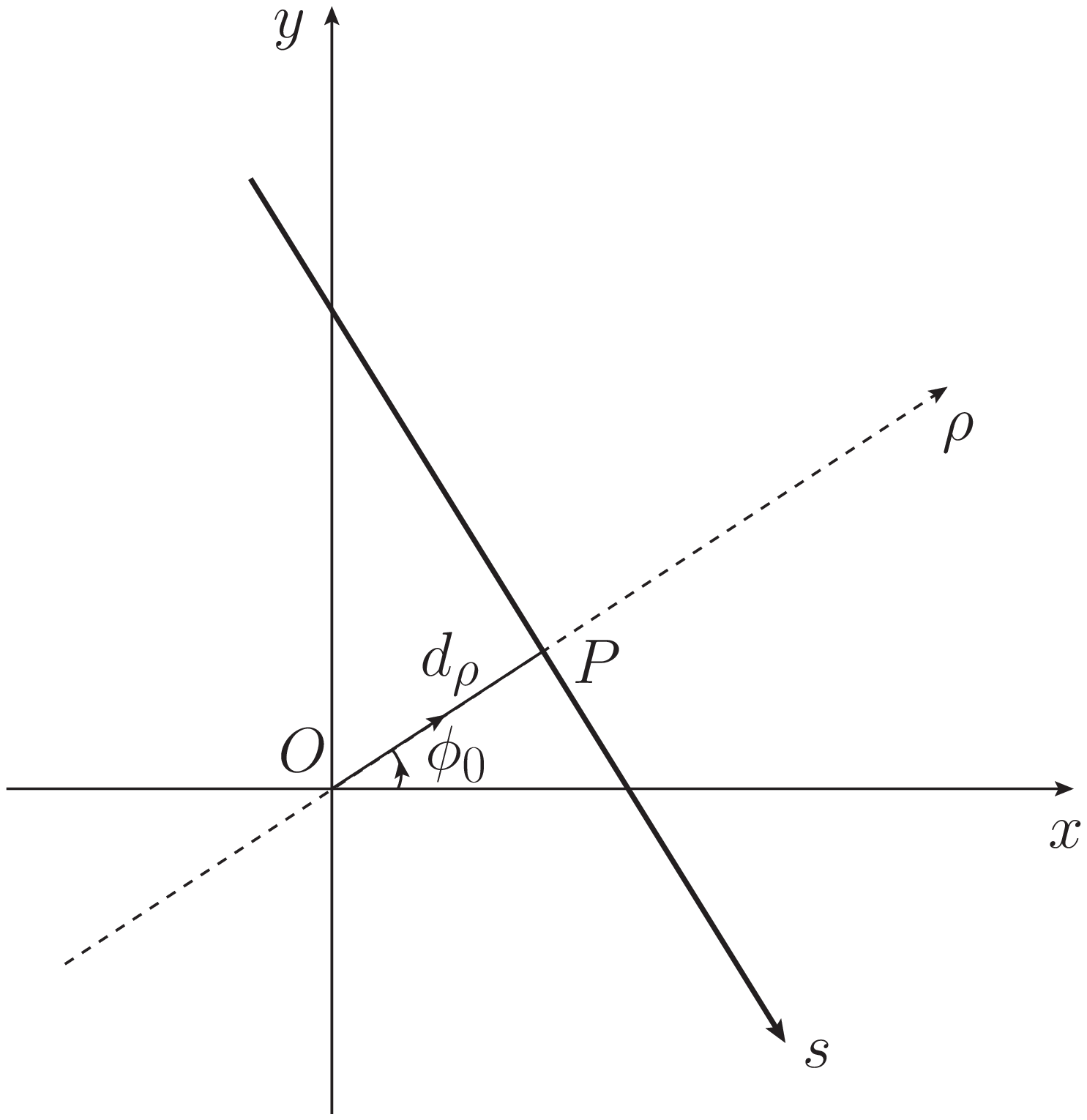}
\includegraphics[width=7cm]{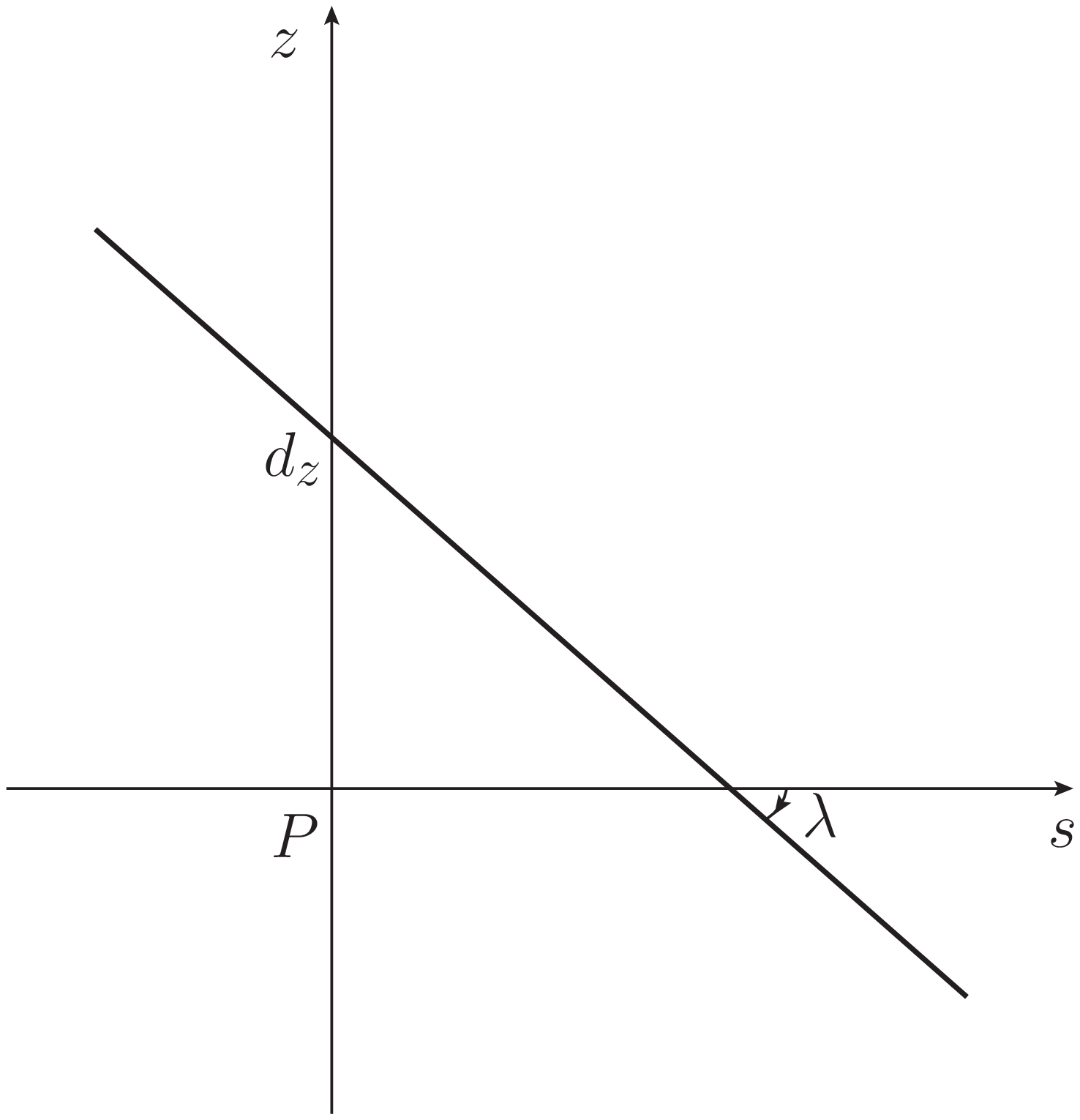}
\caption{Schematic representation of a straight line in the x-y plane (left) and in the s-z plane (right). Left: the projected line in the x-y plane, where
the dashed line $\rho$ is through the origin $O$ and is perpendicular to the line with the point $P$ as the intersection (foot of the perpendicular). The direction $s$ of the projected line in this plane is always defined as downwards. Right: the projected line in the s-z plane, where the intersect is $dz$ and the
angle between the line and the axis $s$ is $\lambda$.}
\label{fig::line_para}
\end{figure}

\begin{figure}
\centering
\begin{overpic}[scale=0.32]{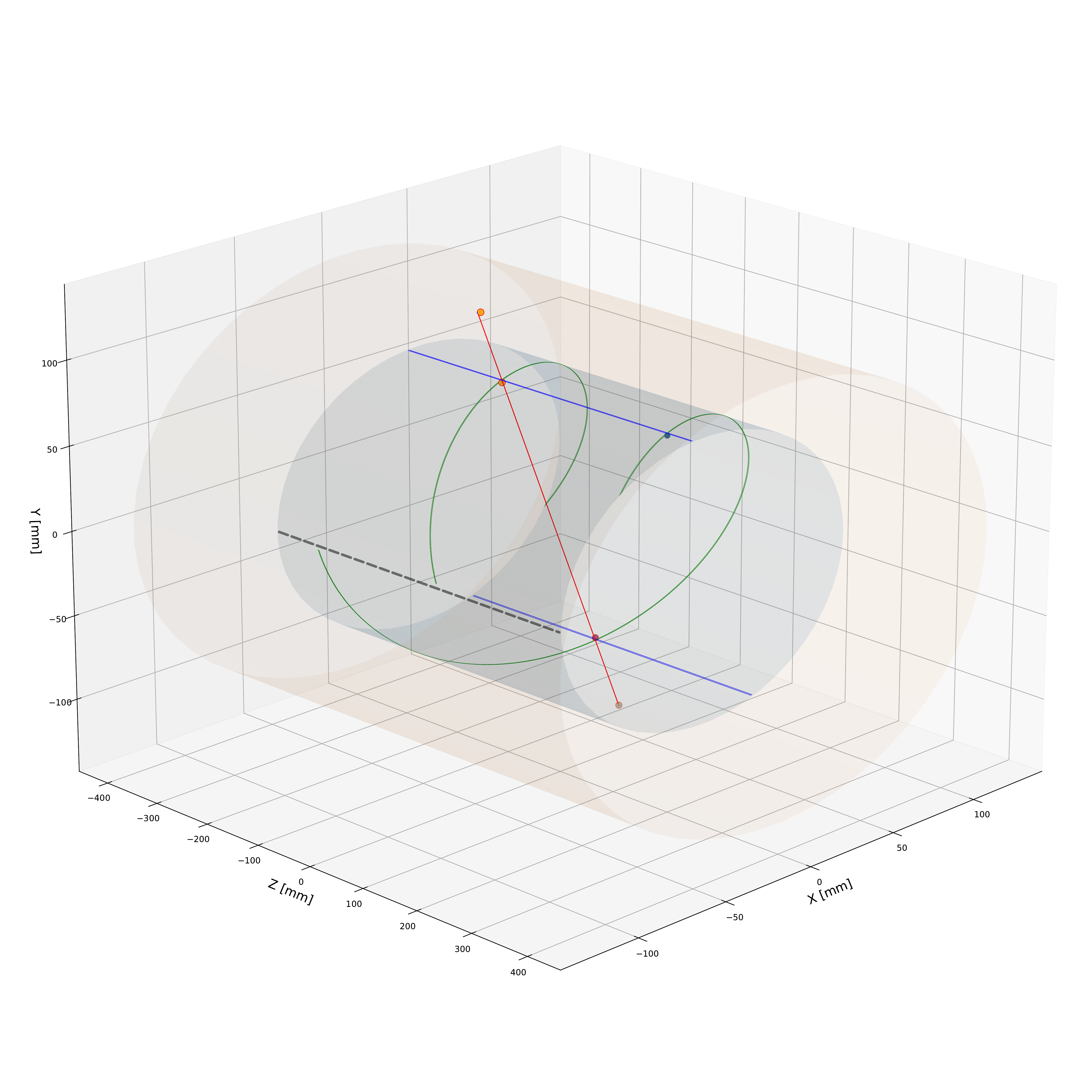}   
\put(65,65){\includegraphics[scale=0.2]{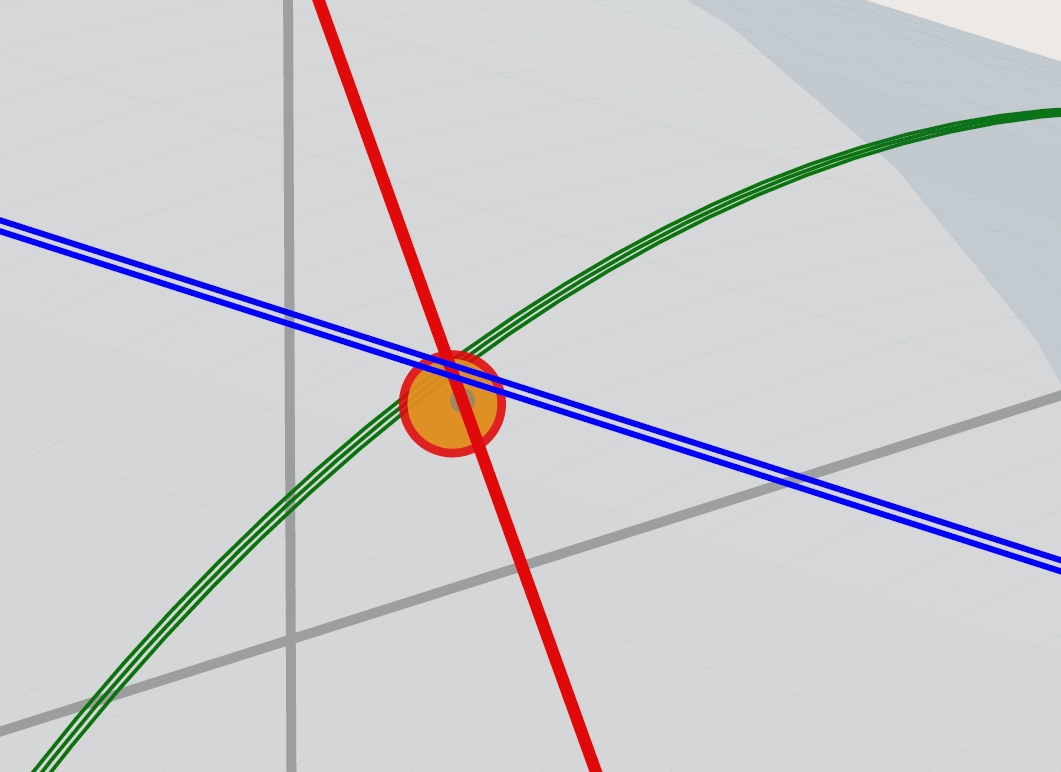}} 
\end{overpic}
\caption{A 3D event display for the 2D cluster reconstruction. The red dots show the collected four 2D cluster when a cosmic-ray pass through the detector. The blue lines show the fired X strips and the green lines refer to the fired V strips.  The dots are the reconstructed 2D clusters. The red line is the fitted track. The top right subplot shows the details of the 2D clusters.}
\label{fig::cluster}
\end{figure}

The trajectory of cosmic-rays is modeled as a straight line and represented by four parameters: $d\rho$, $\phi_0$, $d_z$, and $tan\lambda$ as shown in the Fig.~\ref{fig::line_para}. Here, $d\rho$ is the distance of the line projection from the origin in the x-y plane,  $\phi_0$ is the angle between the perpendicular line $\rho$ and the x-axis in the x-y plane, $d_z$ is the intercept of the line projection in the s-z plane, $tan\lambda$ is the slope of the line projection in the s-z plane. The 2D clusters belonging to one track are fed to the track fitting algorithm, which is based on the least-squares method, to obtain the track parameters. For one cosmic-ray, four 2D clusters should to be collected by the two-layers setup of the CGEM detector as shown in the Fig.~\ref{fig::cluster}.
To reject the fake clusters due to the electronic noise, a simple track-finding algorithm is developed. In the method, at most three clusters with the largest charge on each half of the layer are selected, then all the possible combinations of these four sets of clusters are fitted with the linear model, and at last, the combination with the smallest $\chi^2$ is kept as a track candidate. Figure~\ref{fig::cluster} shows an event display for the track reconstruction.  



\subsection{Alignment algorithm}
\subsubsection{The principle}
To achieve an optimal track parameter resolution, the position and orientation of CGEM layers need to be determined precisely. Based on the cosmic-ray data, the track-based alignment algorithm is implemented~\cite{Blobel:2006yh}. In the track fitting, the track-hit residual is affected by the displacement of the detector, e.g. the positions and orientations of the CGEM differ from the true one, therefore, we can measure the misalignment effect of the detector by fitting a large number of tracks. In the fit, both the track parameters and the misalignment parameters, such as the shift and rotation of the detector to its ideal position, are free. All these parameters are fitted to minimize the deviation between the measurements and track prediction. Here, the track parameters are also named local parameters because they are track-independent. Meanwhile, the parameters to describe the detector displacement are called global parameters because they are shared by all the tracks. The minimized function can be defined as:
\begin{equation}
  \chi^2(p,q)=\sum_{j}\sum_{i} (\frac{m_{ij}-f_{ij}(p,q_j)}{\sigma_{ij}})^2 ,
\end{equation}
where j is the index of track and i is the index of hits in one track. $f_{ij}$ is the track model prediction at the position of the measurement, which is a function of global parameters (p) and local parameters ($q_j$). Usually, $f_{ij}$ should be linearized by Taylor expansion, then the minimization leads to the normal equations $Ca=b$, which is a matrix equation and its dimension is determined by the number of local and global parameters. Here, C is a $(n+m)\times(n+m)$-matrix which is constant for a linear problem, where $n$ and $m$ are the dimensions of global and local parameters. $b$ is the constant part of the first derivative of the $\chi^2(p,q)$ and $a$ is the parameter vector, including both the global and local part, we want to know. The complete matrix equation is given by
\begin{equation}
\left (\begin{array}{c|ccc}
\sum C_{k}^{global} & \cdots & H_{j} & \cdots \\
\hline 
\vdots & \ddots & 0 &0 \\
H_{j}^{T} & 0 & C_{j}^{local} & 0 \\
\vdots & 0 & 0 & \ddots
\end{array} \right )
\times
\left (\begin{array}{c}
	a_{k}^{global} \\
	\hline 
	\vdots \\
	a_{j}^{local} \\
	\vdots
\end{array} \right )
=
\left (\begin{array}{c}
	\sum b_{k}^{global} \\
	\hline 
	\vdots \\
	b_{j}^{local} \\
	\vdots
\end{array} \right ).
\end{equation}
The left side of this equation includes three components. The first part is a contribution of a symmetric matrix $C_{k}^{global}$ with the dimension of the global parameters. All the matrices $C_{k}^{global}$ are added up in the upper left corner of the full matrix of the normal equations. The second contribution is the symmetric matrix $C_{j}^{local}$, which makes a contribution to the full matrix on the diagonal and depends only on the measurement from the j-th track. The third part is a rectangular matrix $H_j$, which has a row number of global parameters dimension and a column number of the dimension of the local parameters.  
To achieve desirable precision,  the needed track number could be in the order of $10^6$. Therefore, the $C^{local}$ will be $t*10^6$ by $t*10^6$ matrix, here, $t$ is the number of the individual track parameters. $H$ will be  $t*10^6$ by n matrix. To be able to solve such a large matrix in a reasonable computing time, a matrix reduction is necessary. We notice that the complete matrix has a special structure, with many vanishing sub-matrices. 
Ignoring the global parameters we could solve the normal equations $C_j a_j = b_j$ for each partial measurement separately by
\begin{equation}
    a_j^{local} = (C_j^{local})^{-1} b_j^{local}.
\end{equation}
The only connection between the local parameters of different partial measurements is given by the sub-matrices $C_k$ and $H$. The special structure of the full matrix allows a matrix reduction such that the complete information from each local track fit is transferred to the global parameter-related matrix. After each track is fitted locally, the corresponding matrices $C_k^{-1}$ and $H_k$ are calculated and added to the global matrix and the vector b in the form of 
\begin{equation}
    C^{\prime}=\sum_{k}C_k^{global}-\sum_{k} H_k (C_k^{local})^{-1} H_k^T,\, b^{\prime} = \sum_k b_k^{global}- \sum_k H_k (C_k^{local})^{-1} b_k^{global}. 
\end{equation}
After the loop over all tracks, the modified normal equations that only contain the global parameters are obtained:
\begin{equation}
    C^{\prime} a^{global} = b^{\prime}.
\end{equation}
The matrix $C^{\prime}$ is reduced to the dimension of the alignment parameters n therefore the computing time is decreased significantly. 

\subsubsection{The implementation at the BESIII}

Since the alignment method is based on the reconstructed track, usually, the alignment procedure will take place after the data taking. To obtain an unbiased and reliable result, both the tracks from cosmic rays and the collision data will be used to determine the displacement between the different parts of the detector. Before the CGEM detector is installed into the BESIII, it will be tested by the cosmic ray.
It is necessary to obtain the displacements between the layers of CGEM detector.  Firstly, the unexpected displacement or distortion could be found in time and fixed before the installation. Secondly, we can obtain the misalignment parameters between the layers and use them as initial values in the following alignment procedure. Therefore, a dedicated alignment study using the cosmic-ray data is performed. 

\begin{figure}[htbp]
\centering
\includegraphics[width=7cm]{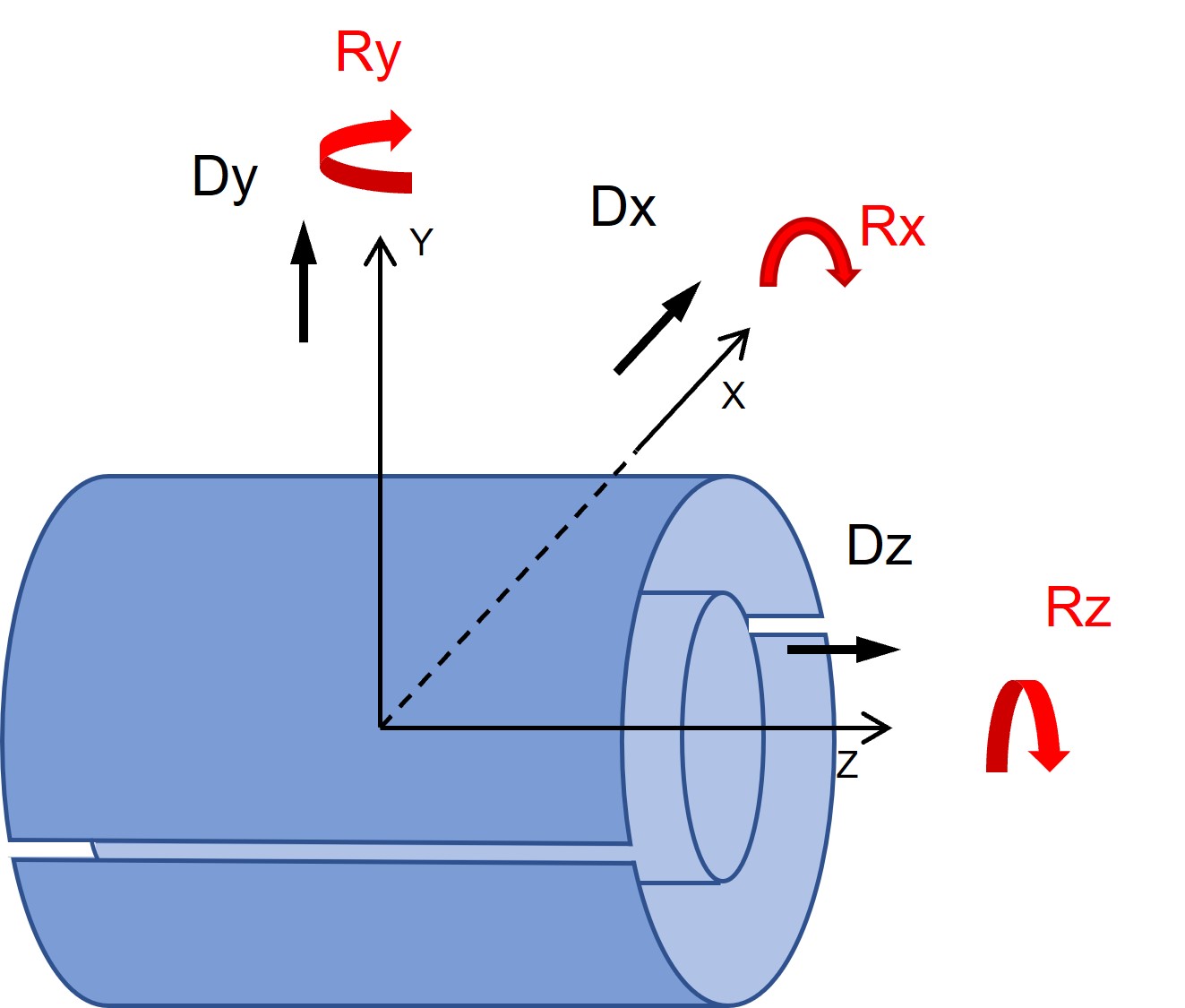}
\includegraphics[width=7cm]{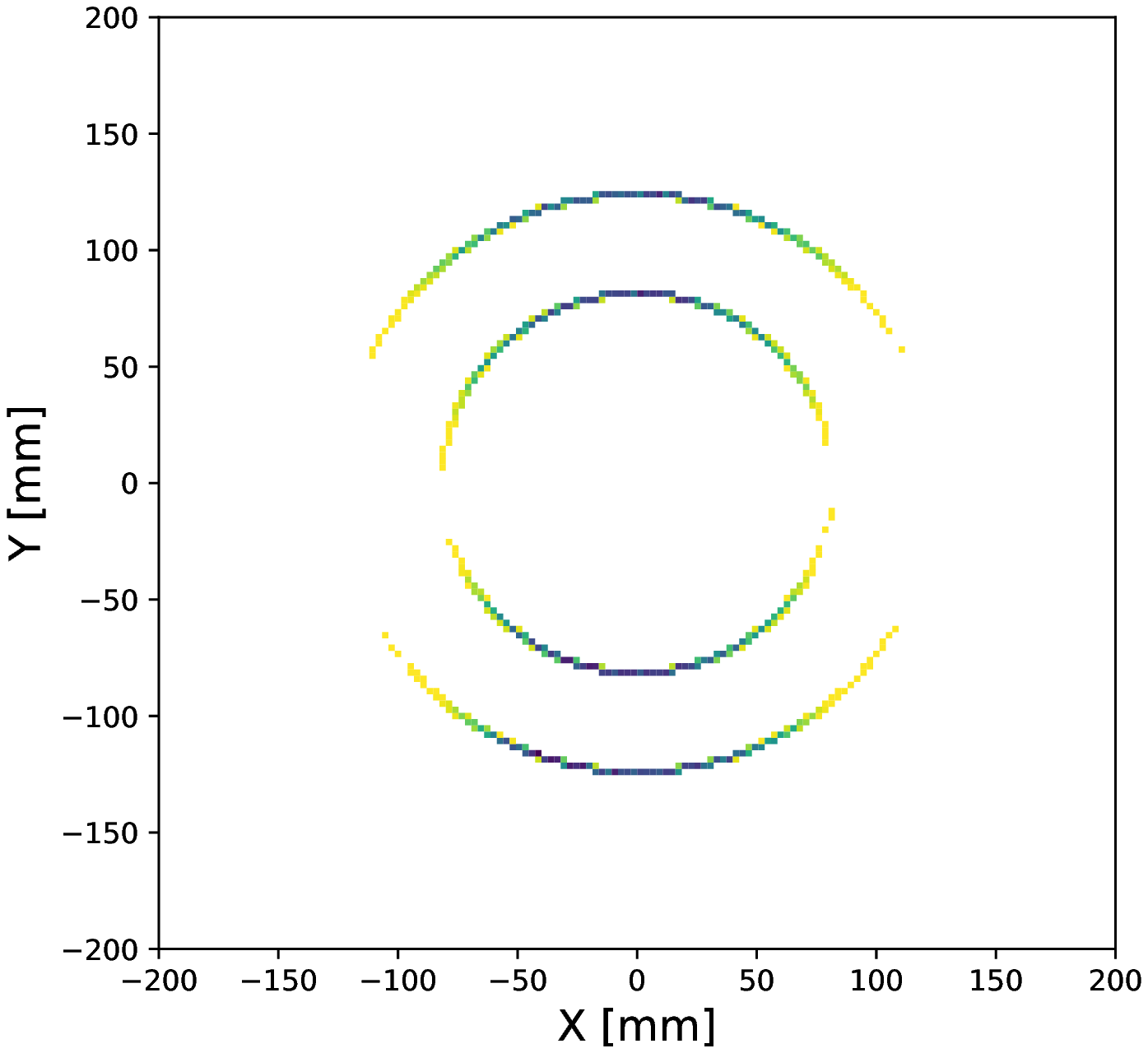}
\caption{Left: the alignment parameters for outer layer. Right: the hit position distribution. }
\label{fig::alignment_para}
\end{figure}

As shown in the Fig.~\ref{fig::alignment_para} left, the cosmic-ray test setup consists of two CGEM layers. The innermost layer is made of one single sheet, we treat it as a rigid body and use it as a reference object to study the displacement of the outer layer. The outer layer includes two sheets covering the top and bottom parts of the detector. So each sheet is considered as an independent component in the alignment study. For each sheet, three translation parameters Dx, Dy, and Dz corresponding to the shifts in the local reference system, and three rotation angles Rx, Ry, and Rz corresponding to small rotations around the respective local axes are taken into account. All these alignment parameters are illustrated in the left plot of Fig.~\ref{fig::alignment_para}.


Several constraints are applied to consider the relations between the alignment parameters. Due to the specific setup for the cosmic-ray data taking, most of the hits are collected in the center part of the detector as shown in the right plot of  Fig.~\ref{fig::alignment_para}. In this case, the Dy parameter is insensitive to the track, therefore, we constrain the Dy values to be zero. In addition, the correlation between Dx and Rz is quite strong since the rotation along the Z axis is equivalent to a shift to opposite directions on the X dimension of the two sheets. To reduce this uncertainty, we require the Dx values of the two sheets to be the same.



\section{Results with the cosmic-ray data}

\subsection{The cosmic-ray data}
The data sets used in the study were collected in 2020. The run numbers are from 10 to 17 and the event numbers are shown in the 
Table~\ref{tab:EventNo}.
The position of the trigger system is slightly different in each run. The setup of the last run, which has the largest statistics, is shown in Fig.~\ref{fig::CosRay_Test} right.


\begin{table*}[htbp]
\begin{center}
\caption{The event numbers in each data set. }
\begin{tabular}{l|cccccccc}
\hline
\hline
Data set      & run10 & run11 & run12 & run13 & run14 & run15 & run16 & run17\\ 
\hline
Event Number  & 18039 & 12831 & 4957 & 2928 & 3318 & 1083 & 5000 & 112405 \\
\hline       
\end{tabular}
\label{tab:EventNo}
\end{center}
\end{table*}

 In the alignment procedure, the initial global parameters are set to zero and the global fit will be performed iteratively until the fit results are converged. During the iteration, the output of the previous fit will be used as the initial parameters of the next iteration.



\subsection{The results}

\begin{figure}[htbp]
\centering
\includegraphics[width=10cm]{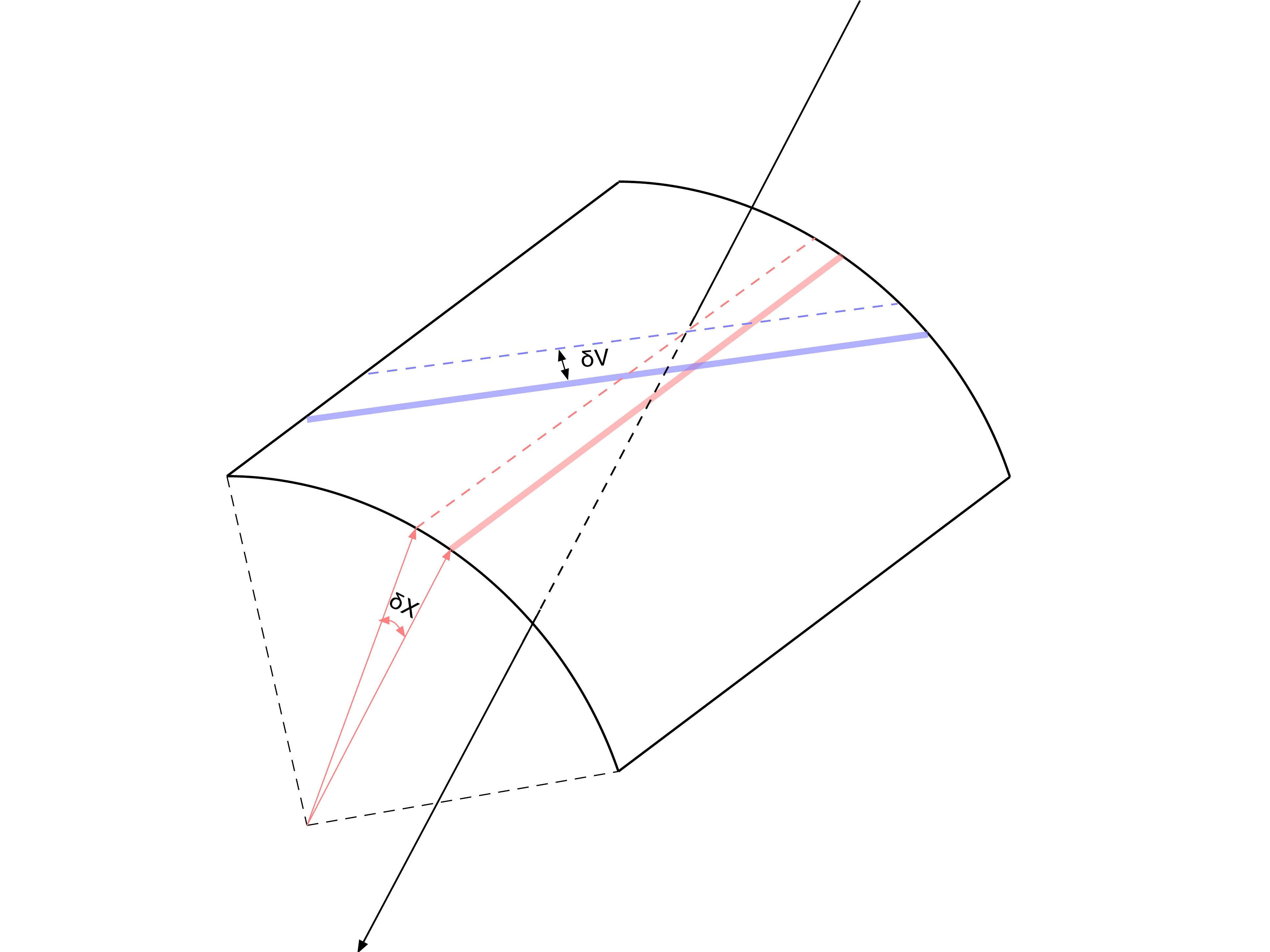}
\caption{The definitions of $\delta X$ and $\delta V$}
\label{fig::dxdv}
\end{figure}

\begin{figure}[htbp]
\centering
\includegraphics[width=\textwidth]{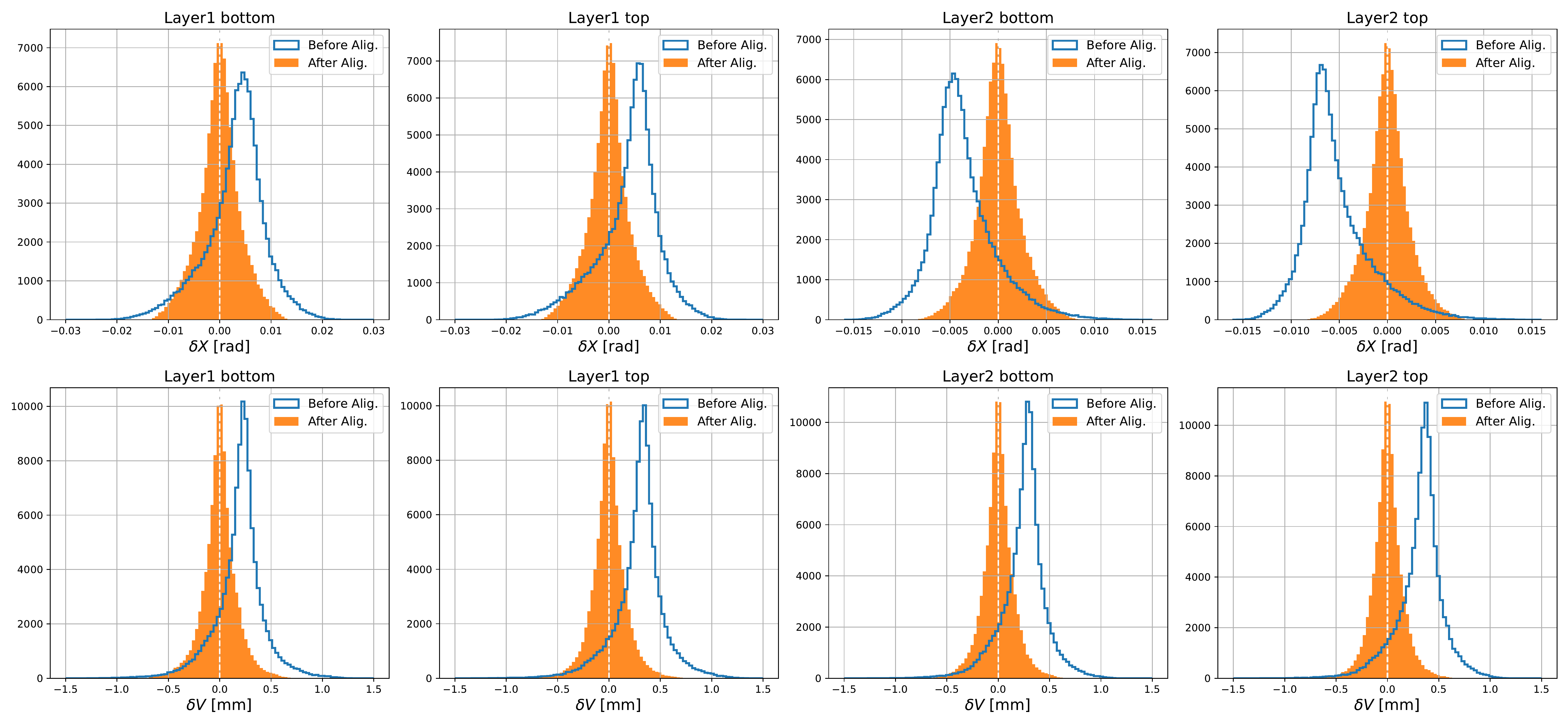}
\caption{The residuals of X and V comparison before and after alignment correction.}
\label{fig::result_dx_dv}
\end{figure}

The inner layer is labeled as "Layer1" and the outer one is labeled as "Layer2". On each layer, the residuals in $X$ and $V$ dimensions (named as $\delta X$ and $\delta V$),  as illustrated in Fig.~\ref{fig::dxdv}, represent the discrepancies between the expected positions and the measurements in the two different dimensions. 
The $\delta X$ and $\delta V$ are investigated before and after alignment correction to check the algorithm as shown in Fig.~\ref{fig::result_dx_dv}. In these plots, clear shifts to zero are found in the residual distributions for both X and V measurements. These shifts due to displacement are corrected after taking the misalignment parameters into account.

\begin{figure}[htbp]
\centering
\includegraphics[width=\textwidth]{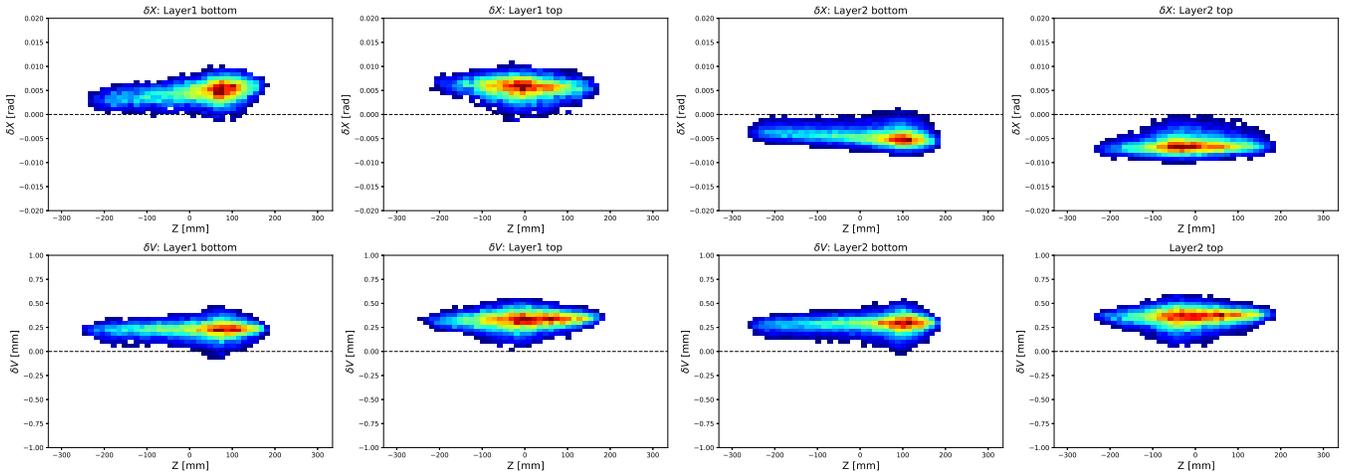}
\caption{Distributions of the residuals as a function of z before alignment correction. }
\label{fig::resi_vs_z_before}
\end{figure}

\begin{figure}[htbp]
\centering
\includegraphics[width=\textwidth]{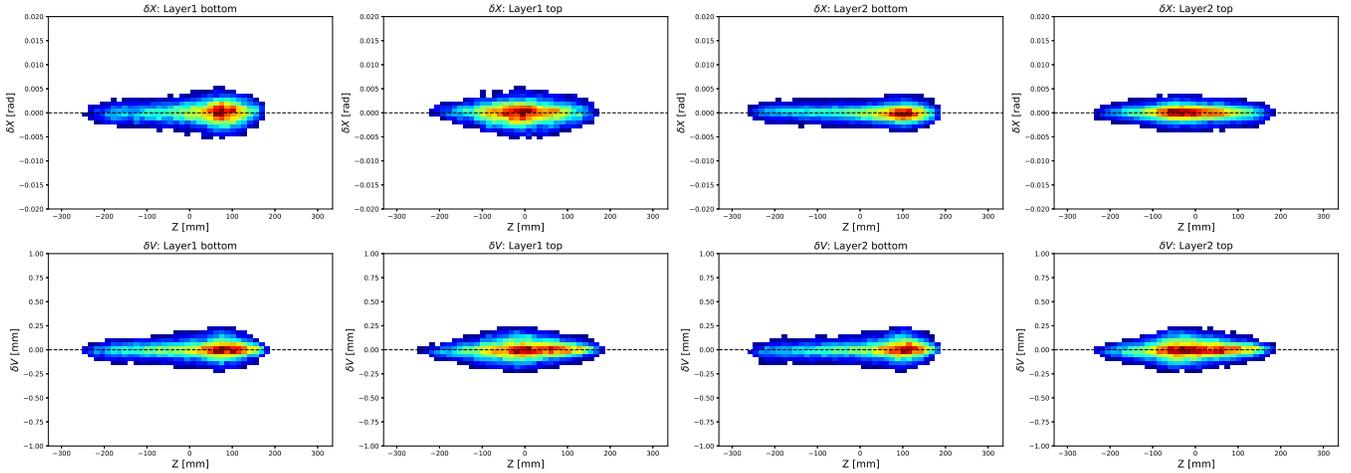}
\caption{Distributions of the residuals as a function of z after alignment correction.}
\label{fig::resi_vs_z_after}
\end{figure}

We also investigate the residual distribution with respect to z. As shown in Fig.~\ref{fig::resi_vs_z_before}, we can find that the $\delta X$s are obviously z dependent.  For example, the $\delta X$ is increased from 3 mrad to 6 mrad in the range of 400 mm along z. This behavior can be interpreted as a rotation of the detector around the Y axis. The scale of the residual and the z-dependent behaviors on the top and bottom part of the same layer are different, which proves that it is necessary to separate the two sheets on the same layer as an individual component in the alignment study. The different layer shows the opposite shift direction and it is also consistent with our expectation. After applying the alignment correction, both the displacement and z-dependent variation are eliminated as shown in Fig.~\ref{fig::resi_vs_z_after}.

\begin{figure}[htbp]
\centering
\includegraphics[width=\textwidth]{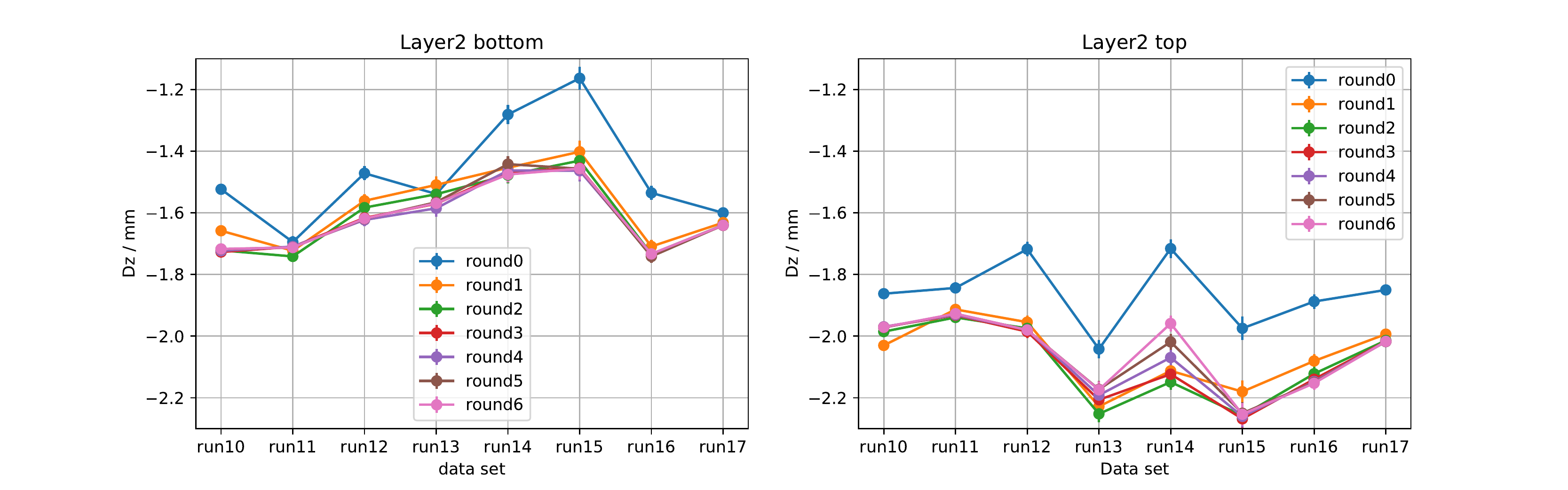}
\caption{The fitted alignment parameter $Dz$ in each iteration from different data sets.}
\label{fig::fitted_dz}
\end{figure}

We check the the fit converges after seven iterations from different data sets and the results are shown in Fig.~\ref{fig::fitted_dz}. We find the fit will converge after 7 times iterations. The fitted global parameters from run10, run11, and run17, which have large statistics, are consistent with each other.
The fitted global parameters from the largest data set run17 are shown in the Table~\ref{tab:final_result}.

\begin{table*}[htbp]
\begin{center}
\caption{The fitted global parameters for the outer layer of the CGEM. }
\resizebox{!}{0.9cm}{
\begin{tabular}{l|ccccc}
\hline
\hline
Position      & Dx(mm) &Dz(mm)&Rx(mrad)&Ry(mrad)&Rz(mrad)\\ 
\hline
Bottom  & $0.377\pm0.008$ & $-1.635\pm0.005$ & $-0.512\pm0.033$ & $0.999\pm0.006$ & $6.528\pm0.064$ \\
Top  & $0.377\pm0.008$ & $-1.984\pm0.005$ & $-2.127\pm0.040$ & $0.302\pm0.007$ & $15.572\pm0.064$ \\
\hline       
\end{tabular}
}
\label{tab:final_result}
\end{center}
\end{table*}

\section{Conclusions and plans}

To summarize, the alignment parameters for the BESIII CGEM detector have been determined using the global fit approach of track-based alignment with the Millepede program. In this study, the cosmic-ray data taken in 2020 with two layers setup is used and about 160000 tracks are fitted simultaneously. Ten alignment parameters are measured with high precision. The results from different data sets are consistent with each other. Clear shift and rotation effects between the two layers are observed. After performing the alignment correction, the consistency between the intersection point determined by the track and the measurement is significantly improved. 

This is the first step of the whole alignment procedure for the CGEM project. The inclusion of cosmic ray tracks in the alignment procedure is not sufficient to control all the weak modes biasing the curvatures of high momentum tracks. After all the three layers will be assembled and installed into the BESIII, further alignment study will be performed using the electron-position collision data together with the cosmic-ray data to get a more precise and reliable result. In addition, the displacement between the CGEM and the outer drift chamber also will be measured with the same software framework.

\section{Acknowledgments}
 We would like to express our great appreciation to the BEPCII accelerator team for their enlightening discussion and helpful suggestions. This work is supported in part by National Natural Science Foundation of China (NSFC) under Contracts Nos. 12275296, 12275297, U1832204, 12175256; National Key R\&D Program of China under Contract No. 2020YFA0406304.





\bibliography{cgem}{}

\end{document}